\documentclass[]{speauth}

\usepackage[utf8]{inputenc}
\usepackage[T1]{fontenc}
\usepackage{lmodern}
\usepackage[english]{babel}
\usepackage{listings}			
\usepackage{subfig}			
\usepackage{tabularx}
\usepackage{longtable}			
\usepackage{colortbl}			

\usepackage{xcolor}     
\usepackage[normalem]{ulem} 

\usepackage{appendix} 

\usepackage{changepage} 

\usepackage[perpage]{footmisc}

\usepackage{multirow}			
\usepackage{array}			
\usepackage[final]{pdfpages}

\usepackage{url}      

\usepackage[bookmarks=false,colorlinks,linkcolor=blue,citecolor=blue,urlcolor=blue]{hyperref}

\newcolumntype{x}[1]{
>{\centering\hspace{0pt}}p{#1}}

\newlength{\longtablewidth}
\usepackage{pdfcomment}

\newcommand{\thistitle}{Abmash: Mashing Up Legacy Web Applications by Automated Imitation of Human Actions}
\newcommand{\thisauthor}{Alper Ortac, Martin Monperrus, Mira Mezini}
\newcommand{\urlaccessed}{accessed Sep. 19 2012}

\author{\thisauthor}

\title{\thistitle}

\date{\scriptsize{Accepted for publication in ``Software: Practice and Experience'' (Wiley) on Oct. 31 2013.}}
\begin{document}

\maketitle

\emph{Abstract:  
Many business web-based applications do not offer applications programming interfaces (APIs) to enable other applications to access their  data and functions  in a programmatic manner. This makes their composition difficult (for instance to synchronize data between two applications). To address this challenge, this paper presents Abmash, an approach to facilitate the integration of such legacy web applications by automatically imitating human interactions with them. By automatically interacting with the graphical user interface (GUI) of web applications, the system supports all forms of integrations including bi-directional interactions and is able to interact with  AJAX-based applications. Furthermore, the integration programs are easy to write since they deal with end-user, visual user-interface elements. The integration code is simple enough to be called a ``mashup''.
}

\section{Introduction}
\label{introduction}

Changes in the application landscape often require to integrate legacy applications with new services, e.g. for improving the information technology (IT) infrastructure, modifying business processes or partnerships, or supporting company acquisitions.
Application integration (a.k.a. EAI for ``Enterprise Application Integration'') enables applications to collaborate, for instance to synchronize some data.
A common integration issue consists in synchronizing the quotation data of a product management application with the one stored in a customer management application.
Analysts state that about a third of the IT budget of enterprises is used for the purpose of application integration \cite{kaib2002enterprise}.

Application integration is facilitated by the availability of application programming interfaces (APIs).
Web applications may provide APIs such as REST or SOAP interfaces.
However, old web applications often lack APIs.

In this case, IT departments face two alternatives for integrating web applications together.
On the one hand, they may dive into the source code of the applications in order to hook integration code at appropriate places. However, this requires a profound knowledge of the structure and the logic of the application, which is often missing due to staff turnover and scarce documentation. 
On the other hand, they can hire people to perform the tasks manually. These tasks usually consist of repetitive tasks like filling in forms or controlling system parameters by using a web interface. However, the results are error-prone, because it is unlikely for humans to do flawless and accurate work over large amounts of data. Interactions can easily be mixed up, for example by interchanging two date fields. 

This paper presents a system that aims at facilitating the integration of web applications that do not offer programming interfaces. 
For ease of reference, we call them in the following simply \emph{legacy web applications}.

\subsection{Design Requirements}
\label{sec:req}

We aim at creating a software system that supports the integration of legacy web applications. 
By integration we mean collaborations between two or more applications: migration of data, collaboration in a process orchestration, etc.
This section presents the requirements guiding the design of such a system and their rationales.

\medskip
\emph{R1: The system shall support the integration of web applications that do not offer application programming interfaces.}\\
\emph{Rationale:} Many web applications have been designed and developed without an API. Improvements of the IT infrastructure and business process evolutions require either integrating those legacy applications or re-implementing them. The latter being very costly \cite{kaib2002enterprise}, the former is key for cost-effectiveness. 

\medskip
\emph{R2: The system shall support all forms of integrations including bi-directional interactions.} \\
\emph{Rationale:} Many kinds of application integration require bi-directional interactions.
For instance, one may need to synchronize the quotation data of a product management application with the one stored in a customer management application. 
In this synchronization example, data is read and submitted on \emph{both} side (bi-directional), so that the data of the product management application corresponds to the data of the customer management application.
Many mashups approaches read data from various sources but do not submit data into third-party applications.

\medskip
\emph{R3: The system shall support all kinds of HTML based web applications including form-based and AJAX-based applications.\footnote{Web technologies such as Silverlight or Flash apps are out of the scope of this paper.}}\\
\emph{Rationale:} While pure HTML web applications were the norm in the early web, there are now so-called AJAX applications where the client browser executes a large amount of code (usually Javascript). In a realistic setup, an integration program interacts with both kinds of applications, for instance reading data from a form-based application and submitting this data into a AJAX-based web user-interface or vice versa.

\medskip
\emph{R4: Integration programs should be easy to write.} \\
\emph{Rationale:} Ideally, integration programs are jointly written by the experts of the applications to be integrated. In reality, the experts may have moved on to another project or to another company, the company maintaining a key software package may have sunk, and there may only be some scarce documentation.
That is to say, to our knowledge, most integration programs are to be written by non-expert developers.
While this requirement is hardly testable, it is nonetheless very important: if an integration program costs $X$ to be set up, a system enabling companies to pay  $X/2$ for the same program is better. 
Another way to formulate this point is, the less the knowledge required for writing an integration program, the better for companies.

\subsection{Research Challenge}

To the best of our knowledge and as we elaborate in the related 
work section (cf. section \ref{related_work}), there is no approach that satisfies the aforementioned requirements all together. 
Conventional approaches to integrating legacy code based on classical middleware fail
to satisfy {\bf R1}, {\bf R4}, as they imply a deep understanding 
 of the application source code and database schemas in order to invasively modify them  \cite{Vinoski2003}. 
Approaches that integrate legacy software into a service-oriented architecture by automatically wrapping 
PL/I, COBOL and  C/C++ interfaces (e.g., \cite{Sneed2006}) into SOAP-based web services are not suited for the domain
at hand, as they do not target web applications with no API ({\bf R1}) and do not provide bi-directional integration mechanisms ({\bf R2}). 
Declarative web query languages such as CSS selectors are powerful, but require much more tooling with respect to working with client-side javascript code ({\bf R3}).
Many modern mashup approaches, such as Yahoo Pipes! \cite{fagan2007mashing}, focus on easily composing  different sources together by short scripts, they are not 
suitable either, as they make the strong assumption that a programmable interface or a web service exists, which is not true for legacy web application ({\bf R1}) \cite{liu2007towards}.

Hence the research challenge addressed in this paper can be formulated as \textbf{how to integrate legacy web applications with little knowledge of the code of the involved applications?}

\subsection{Contribution}

In this paper, we take another perspective on the integration of web applications. 
We propose an approach, called Abmash, that supports content extraction and web interface automation
by emulating, imitating sequences of human interactions in a programmatic manner.
We call integration programs written with our approach \emph{integration mashups}, or simply (Abmash) 
mashups, when there is no risk for ambiguity, to highlight their affinity to end-user mashups in terms of  
easiness of developing them (but there not end-user mashups).

We evaluate our approach in three different ways.
First, we provide an in-depth discussion of prototype integration mashups implemented with our approach.
Second, we present a user study that aims at evaluating the degree of easiness of learning and 
using Abmash to create integration mashups.
Third, we thoroughly compare Abmash with other alternatives.
Those evaluations indicate that Abmash is a solid foundation for integrating and composing legacy 
web applications together. The first author of this paper also uses Abmash in his own business.

\section{Overview of Abmash}
\label{overview}

This section presents an overview of Abmash. We first define the key concepts behind our contribution and then present the vision in short, before giving an overview on how it works. The section also contains links to the related sections for selective reading of the paper.

\subsection{Important Definitions}
Let us first define the main concepts related to our contribution.
\emph{Application integration} software is software that \emph{``interconnects disparate systems to meet the needs of business'}, in order to \emph{``ease the pain of integration''}, \cite{Linthicum2000}.
As stated above, we specifically address the integration of legacy web applications with no programming interfaces.
As discussed later in the related work section, there is no unique definition of ``mashup''. In this paper, we take a large-scope definition inspired from \cite{wood2007interactive};  a mashup is \emph{``the integration of available applications using web-based technologies''}.
Finally, the \emph{visual semantics} of web pages consists of \emph{``spatial and visual cues [that] help the user to unconsciously divide the web page into several semantic parts''} (\cite{deng2003vips}).
Section \ref{related_work} extensively discusses those research fields.

\subsection{What is Abmash?}

Inspired by the work by Cai et al. \cite{deng2003vips}, the key idea behind Abmash is to solve integration problems by using the visual semantics of web pages.
Abmash is a framework that enables integration code to imitate sequences of human interactions in a programmatic manner.
In other terms, integration is done at the presentation layer, everything via the user interface of legacy web applications.

To give the reader a first intuition on what we mean, we simulate the process of adding a product in a database. From the human-computer interface, this activity 
consists of typing the textual description in a field called ``Description'' and clicking on a button labeled ``Add''. 
Using the programming interface of the Abmash framework proposed in this paper, this sequence can be programmatically imitated by the 
following code snippet (Section \ref{framework_description} describes the Abmash programming interface);

\begin{verbatim}
... Java code ...
browser.type("Description","Blue Shoes");
browser.click("Add");
... Java code ...
\end{verbatim}

From a conceptual point of view, Abmash addresses the requirements aforementioned with 1) the idea of using the visual semantics of web pages and 2) an API that allows developers to write short and self-described integration programs (this kind of API is sometimes referred to as \emph{fluent API}).

The benefits of the proposed approach are twofold.
First, the difficulty of writing the integration code is much alleviated: the integration engineer has to understand none of:
1) the legacy application code;
2) the legacy database schema and constraints;
3) the produced HTML code, i.e. there is no need for comprehending the document object model in terms of tags (e.g. HTML's \texttt{<div>}, \texttt{<ul>}) or attributes (e.g. identifiers, CSS classes) \cite{gupta2003dom}.
Second, since the integration is done at the level of the end-user interface, it benefits from the  logic 
related to business functions and integrity checks that prevents users to enter incorrect or partial data.
The evaluation section (\ref{evaluation}) deepens those points.

\subsection{How it Works in a Nutshell}

An Abmash integration program is written in plain Java. 
It uses classes and methods of the framework to find elements on the web page (size, position, color, etc.) and to execute user interactions on these elements (\texttt{type, click, etc.}).
These framework classes and methods are implemented by calling and remote controlling a Firefox browser that is responsible for computing 
all the visual rendering of web pages and client-side execution of JavaScript.
For doing so, Abmash uses an intermediate software component, 
called Selenium\footnote{\url{http://seleniumhq.org/} \urlaccessed}, which wraps up some of the browser machinery into a regular yet low level programming interface. Section \ref{application_architecture} describes which parts of Selenium is used by Abmash and Section \ref{comparative_study} the added value (ease of use and conciseness).
For a developer to write an Abmash integration program, 
she only has to know the 
programming interface of Abmash.
For debugging purposes, developers can watch the execution of their Abmash applications, 
which gives an immediate visual feedback about the correctness of the visual queries and interactions.

In other terms, Abmash hides the complexity of both (a) the applications to be integrated 
and (b) the technical details to query and manipulate elements based on visual semantics. 
Section \ref{application_architecture} gives more details on the implementation of our approach.

\section{Description of the Abmash Framework}
\label{framework_description}

This section presents a framework, called 
Abmash\footnote{The latest API Documentation of Abmash can be found at \url{http://alp82.github.com/abmash/doc} \urlaccessed.}, for integrating legacy web applications
with no programming interface (as characterized in the introduction).
Abmash offers an API for writing mashups to compose such legacy web applications  by imitating human interactions with their visual interface.

\subsection{How to Use the Framework?}

To use the framework, a developer has to 1) understand the concept of ``programming with the visual semantics''
and 2) get familiarized with the Abmash API.
The expected users of Abmash are not end-users, they are Java developers responsible on integrating legacy applications with no API, but with little or no knowledge of the code of the applications to be integrated together.

For the former, a developer should put aside everything she knows about web services and queries on document object models, if she has such a background.
At heart, an Abmash program is a natural language thought like: \emph{go to Google, type ``Abmash'' in the search field, submit the query and get the \#1 result (the first link that is below the search field)}. By reading the API, and browsing code snippets, the programmer learns that the main class of Abmash is ``Browser''\footnote{
Creating a \texttt{Browser} instance simulates launching a browser. \texttt{Browser} has many methods to support fluent programming. We have already seen the \texttt{query} method. Other methods include \texttt{browser.openUrl(url)} to navigate to the desired URL, \texttt{browser.history().back()} and \texttt{browser.history().forward()} to navigate through the browser history.
}
.
Within 45 minutes (an order of magnitude taken from our user experiment, see \ref{user_experiment}), she would come up with code similar to Listing \ref{fd_first_example}.
This program imitates human understanding of the visual appearance of web pages and human interactions with a web application. 
Note that no knowledge of the underlying HTML or CSS is involved at all\footnote{Our framework is implemented in Java, hence the listings of this section will use this programming language. However, the API should be implementable in any language that supports some kind of fluent interface.}

\begin{lstlisting}[caption={Exemplary Abmash application to perform a Google search and to extract content. No knowledge of the underlying HTML or CSS is involved.},label=fd_first_example,float]
// open a browser instance and navigate to Google
Browser browser = new Browser("http://www.google.com");

// enter search term into input field, then submit
HtmlElement searchField = browser.type("search", "Abmash").submit();

// get the title of the first result based on the visual rendering of the results
// the query consists of three predicates which must be true to return a matching result
String title = browser.query(headline(), link(), below(searchField)).findFirst()
                       .getText(); 
\end{lstlisting}

The framework consists of classes and methods that simulate human understanding and interactions.
They are grouped into different concerns.
First, one can select elements in web pages with \emph{visual queries} (Sec. \ref{visual_selection}).
Further, the framework enables developers to imitate  \emph{mouse- and keyboard-based interactions} (Sec. \ref{mouse-based}).
Finally, we will also discuss how AJAX interactions are supported (Sec. \ref{framework_javascript}).

\subsection{A Fluent Programming Interface}
A key design decision is that we chose a fluent interface for programming with the Abmash framework.
An API has a fluent programming interface \cite{Fowler2005} if the program reads similarly to natural language.
However, complete fluency is not always possible, certain part of the framework are more fluent than others.
Fluent interfaces are usually associated with object-oriented programming, where method calls can be chained naturally, and the chaining of method names can resemble to natural language if designed as such.
As shown in Listing \ref{fd_first_example}, the code queries the browser for headline links that are below the search field and then extracts the text of the first one: the code is somewhat close to this natural language description.

\subsection{Visual Queries}
\label{visual_selection}

To ease and accelerate writing queries over the HTML structure of web applications, Abmash supports to express these 
queries based on visual properties of web pages, such as visible text and element locations on the screen (rather than in the source).
For instance, listing \ref{fd_has} shows an Abmash query to select all elements containing the text ``price''.
Queries can have an arbitrarily nested amount of predicates, thanks to the fluent interface.
All predicates are case-insensitive. In the following, we describe all types of predicates that Abmash provides.

\subsubsection{Contains Predicate}
The \texttt{contains} predicate checks whether an element contains the specified text. 
First, priority is given to elements whose visible text matches exactly or at least partially (e.g., the query word may be part of the visible text). 
If not found, this predicate also checks whether the specified string is contained in HTML attributes such as ``name'', ``title'', ``class'' and ``id'' (the order of priority depends on the element type). 
The rationale is that developers tend to use semantic identifiers in HTML, e.g. the HTML element coded 
as \verb|<div id="price">| is likely to contain a price.
The  \texttt{contains} predicate always returns the most specific HTML elements (i.e., the deepest in the DOM tree and not their parents).
Listing \ref{fd_has} shows a short example of using the \texttt{contains} predicate.
\begin{lstlisting}[caption={Filter by visible or attribute text},label=fd_has,float]
// Find all elements containing "price"
HtmlElements priceElements = browser.query(contains("price")).find();
\end{lstlisting}

\subsubsection{Element Type Predicates}
\label{framework_element_type_predicates}
On screen, human users can recognize the following kinds of elements:
\begin{enumerate}
\item \textit{text()}: elements with text in it
\item \textit{headline()}: titles and other elements with bigger font size
\item \textit{clickable()}: clickable elements like links or buttons
\item \textit{typable()}: input elements which can be used to enter text
\item \textit{checkable()}: input elements like checkboxes or radioboxes
\item \textit{choosable()}: input elements like drop downs or selectable lists
\item \textit{datepicker()}: input elements which allow the user to select a date and time
\item \textit{submittable()}: input elements to submit a form
\item \textit{image()}: image elements
\item \textit{list()}: list elements
\item \textit{table()}: table elements
\end{enumerate}

Abmash selectors use these high-level types rather than low-level HTML tags which would require to browse the HTML source code. For instance, listing \ref{fd_istitle} shows how to select all headers of an HTML document. 
These methods have an optional text argument which may further specify what to search for as visual text. 
If there is, for example, a select box for the display language of a web application, the element can be simply found by searching for choosable items containing the language name (see listing \ref{fd_is_choosable}).
In the following, we comprehensively describe those kinds of web elements.

\textit{Texts} are elements that contain at least one character of visible text. The \texttt{text()} predicate can be used to find elements with specific visible text if a text string is given.

\textit{Headlines} are indicated by the use of the header elements \texttt{<h1>}, \texttt{<h2>}, \texttt{<h3>}, \texttt{<h4>}, \texttt{<h5>} and \texttt{<h6>}. Since not every web page uses semantic markup to structure their content, \texttt{headline()} also selects elements which have a font size larger than the default size on that page. This definition closely corresponds to the visual representation of the web application.

Defining a \textit{clickable} element is not trivial. Besides links (\verb|<a href="...">|) and some form elements, 
every element on a page can be clicked on if a click event handler is attached to this element with JavaScript. 
The \textit{clickable()} selector selects links (\texttt{<a href="\ldots">} tags), buttons (\texttt{<button>}, \texttt{<input type=``button''>}, \texttt{<input type=``submit''>}) and form elements (i.e. checkboxes, radioboxes, etc.). 
It then adds in the set of clickable elements those for which a click listener has been attached.
\texttt{clickable()} is often combined with a \texttt{contains()} predicate to select clickable elements that contain attribute values or visible text with the specified string.

\textit{Typable} elements are form fields that are designated for text input. Additionally, there are WYSIWYG text editors\footnote{What You See Is What You Get, e.g. TinyMCE see \url{http://www.tinymce.com/} \urlaccessed}, which can be loaded into the document by using JavaScript code. 
Abmash also considers those elements as typable.
Labels of a typable input elements are usually visually close to each other, but may be difficult to associate when looking at the source code. Abmash automatically searches for typable elements with a label nearby if a text string is specified (similar to the clickable predicate).

\textit{Checkable} elements are checkboxes or radioboxes in forms. 

\textit{Choosable} elements are selection lists or dropdown boxes in forms. 

\begin{lstlisting}[caption={Find choosable element by visible text},label=fd_is_choosable,float]
// Find language dropdown box by its field label
HtmlElement languageBox = browser.query(choosable("language")).findFirst();

// Alternative: find language dropdown box by specific option value
HtmlElement languageBox = browser.query(choosable("spanish")).findFirst();
\end{lstlisting}

\textit{Datepicker} elements are form elements that allow the choice of a specific date and time. The \texttt{datepicker()} method identifies those elements.

\textit{Submittable} elements are form elements to submit the corresponding form. The \texttt{submittable()} method identifies those elements.

\textit{Image} elements are identified by \texttt{<img>} tags. Further, all elements with the CSS attribute \texttt{background-image} are selected.

Content is often structured in lists or tables to visualize the interconnections between the provided information. The predicates \texttt{list()} and \texttt{table()} select such elements if they use the appropriate markup, namely \texttt{<table>} and \texttt{<ul>}, \texttt{<ol>} or \texttt{<dl>}. Furthermore, the interaction with specific table cells is simplified by the use of the \texttt{Table} class. It provides methods to select specific cells depending on their corresponding row and cell descriptions. Listing \ref{fd_is_table} shows a concrete example. Note that the \texttt{getCell} method is very close to what the developers sees on screen (e.g. the cell in the second row and the fourth column, or the cell in the third row and in the column labeled "Mail").

\begin{lstlisting}[caption={Selecting elements based on high-level element types},label=fd_istitle,float]
// Find all title elements
HtmlElements titles = browser.query(headline()).find();

// Find all title elements which are clickable
HtmlElements clickableTitles = browser.query(headline(), clickable()).find();
\end{lstlisting}

\begin{lstlisting}[caption={Advanced table handling},label=fd_is_table,float]
// search for a table which contains the query string "Username"
// and returns the internal table representation of it
Table userTable = browser.getTable("Username");

// return the cell in the second row and the
// fourth column (index numbering is starting at 0)
String cellText1  cell = userTable.getCell(1, 3);

// equivalent method that is more like human understanding of web pages
// return the cell in the third row and the
// column labeled "Mail"
String cellText2  = userTable.getCell(2, "Mail");

\end{lstlisting}

\subsubsection{Visual Predicate}
\label{framework_closeness}
The aforementioned predicates often are not sufficient for selecting the desired target elements. For example, selecting the user images (avatars) of bulletin board threads is difficult because they can not easily distinguished from other images on the same page. With Abmash, elements visually close to another reference element can be selected depending on their distance from each other. As shown in listing \ref{fd_closeness}, it is also possible to define a direction, so that elements that are located in another direction are filtered out. Possible directions are ``below'', ``above'', ``leftOf'' and ``rightOf''.

\begin{lstlisting}[caption={Visual closeness and direction},label=fd_closeness,float]
// Find all usernames and get the avatar images below them
HtmlElements avatars = browser.query(image(), below(contains("username"))).find();
\end{lstlisting}

Visual predicates use the rendered visual output of the web page. The alignment of elements is defined by their x and y coordinates and by their width and height.
Each visual predicate in a specific direction has one of the three following different forms: 

\begin{itemize}
\item Closeness to a single specified reference element in the specified direction (for example \texttt{below()})
\item Closeness to a set of specified reference elements in a specific direction, while the element has to match the criteria for any reference element  (for example \texttt{belowAny()})
\item Closeness to a set of specified reference elements in a specific direction, while the element has to match the criteria for all reference elements  (for example \texttt{belowAll()})
\end{itemize}

All distances are calculated between the centers of each source-destination pair with the Euclidean metric.

\subsubsection{Color Predicate}
\label{color_predicate}
Humans easily recognize and describe web page elements by their color. Therefore, Abmash allows the developer to select elements with specific colors. Color queries consist of three parameters:
\begin{itemize}
\item the color name or RGB value
\item the tolerance of the query, higher values include similar colors
\item the dominance of the color, lower values return elements which contain additional other colors
\end{itemize}
An example usage of the color predicate is shown in listing \ref{fd_color}.

\begin{lstlisting}[caption={Color filter},label=fd_color,float]
// Find all blue images with default tolerance and dominance
HtmlElements blueImages = browser.query(image(), color("blue")).find();

// custom tolerance and dominance
// high tolerance: in order to be selected the element color has to be less similar to "blue" in comparison to the default setting
// low dominance: in order to be selected the element may contain more other colors in comparison to the default setting
HtmlElements blueishImages = browser.query(image(), color("blue", Tolerance.HIGH, Dominance.LOW)).find();
\end{lstlisting}

\subsubsection{Boolean Predicate}
It has been shown that an arbitrary amount of predicates can be chained to narrow down the result set (see an example in listing \ref{fd_chain}).
\begin{lstlisting}[caption={Chained filters},label=fd_chain,float]
// a combination of multiple predicates to narrow down the result:
// this query selects clickable images in red below elements with the text "summary"
HtmlElements elements = browser.query(clickable(), image(), color("red"), below("summary")).find();
\end{lstlisting}

In addition, Abmash also provides developers with predicates containing boolean operators. With logical \textit{OR} predicates, the result set can be extended, logical \textit{AND} predicates are used to narrow down the resulting elements (this is the default behavior) and logical \textit{NOT} predicates invert the result.
Boolean combinations are fully nestable (no restriction on the depth) and therefore allow formulating very complex queries.
An example is shown in listing \ref{fd_sub}.

\begin{lstlisting}[caption={Using boolean predicates},label=fd_sub,float]
// select all elements that contain the texts "car" or "motorcycle"
// but exclude clickble as well as typable elements from the result set
HtmlElements elements = browser.query(
	or(contains("car"), contains("motorcycle")),
	not(or(clickable(), typable()))
).find();
\end{lstlisting}

\subsubsection{Backwards Compatibility}
\label{shortcuts}
Experienced programmers may want to use usual selectors (XPath, CSS or jQuery). This is possible by using the following two predicates alongside other visual predicates:
\begin{enumerate}
\item \texttt{selector(selector)} with selector being a CSS or jQuery selector
\item \texttt{xpath(selector)} with selector being an XPath selector
\end{enumerate}

\subsubsection{Ordering the Result}
\label{ordering_result}
When a query is executed, Abmash retrieves a set of HTML elements. Often, the programmer takes the first one, assuming that is the best.
Abmash uses a number of heuristics to optimize the order of elements returned by a query:
\begin{itemize}
\item elements returned by a visual predicate are ordered by the smallest visual distance
\item elements returned by a color predicate are ordered by the closest color distance with respect to the tolerance and dominance settings
\item elements with a matching label nearby have higher priority because they are likely the desired target
\item elements with a smaller size are given a higher priority because they are considered more specific
\end{itemize}

For queries that contain text, the elements are weighted with the following predicates, with descending order (exemplary use of the query string ``user''):
\begin{enumerate}
\item \textbf{exact} matches, i.e. \texttt{<div>User</div>} or \texttt{<div class="user">John</div>}
\item \textbf{word} matches, i.e. \texttt{<p>The following user is awarded a present.</p>}
\item \textbf{starts-with} or \textbf{ends-with} matches, i.e. \texttt{<div>Username</div>} or\\
\texttt{<span class="mail-of-user">foo@bar.com</span>}
\item \textbf{contains} matches, i.e. \texttt{<label>Superusers</label>}
\end{enumerate}

\subsection{Mouse and Keyboard Interactions}
\label{mouse-based}
Mouse interaction includes clicking on and hovering over page elements. 
Listing \ref{fd_mouse} demonstrates Abmash constructs to model clicking on single or multiple elements.
\begin{lstlisting}[caption={Mouse interaction},label=fd_mouse,float]
// click on all "Publish" elements, e.g. checkboxes
browser.query().isClickable().has("Publish").find().click();

// shortcut method that fetches the same elements and selects the first one to click on
browser.click("Publish");
\end{lstlisting}

Drag and drop commands can be performed by using the \texttt{browser.drag(element\-To\-Drag, elemenToDropOn)} method. In addition, the target element can be defined using absolute or relative coordinates by calling the \texttt{browser.drag(elementToDrag, x, y)} and \texttt{browser.dragBy(element\-ToDrag, x, y)} methods. The first method drags the source element to the specified position, the second method drags it relatively to its current position.
Double clicks are performed with the \texttt{element.double\-Click()} method and right mouse button clicks with \texttt{element.rightClick()}. The \texttt{element.hover()} is used to move the mouse cursor on top of the specified page element without clicking it.

\label{keyboard-based}

Similarly to mouse clicks, key presses are executed on target elements, which are found with the given query string. 
This is mainly achieved with the methods \texttt{type()} and \texttt{keyPress()}.  
Listing \ref{fd_keyboard} illustrates the different usage possibilities.The use of combinations to enter text and press keys covers all possible keyboard interactions. Shortcuts as well as key sequences can also be performed.
Additionally, keyboard interactions can be arbitrarily combined with mouse interactions.
\begin{lstlisting}[caption={Keyboard interaction},label=fd_keyboard,float]
// get first returned element (an input field with the label "Description") and enter the text
browser.type("description", "Text which will be entered in the input field.");

// simulating key strokes
// the first keypress searches for the element labled "search", hits the right arrow key
// and returns that element, which is then used to press the Enter key
browser.keyPress("search", "right").keyPress("enter");

// combining keyboard and mouse interactions on an
// exemplary selection list which supports multiple choices
browser.keyHold("ctrl");
browser.click("salt");
browser.click("onion");
browser.click("vinegar");
browser.keyRelease("ctrl");
\end{lstlisting}

In the following is a summary of all methods which are offered by the \texttt{Browser} class to simplify the execution of often used interactions. Most of these methods were already presented in the preceding code examples:
\begin{itemize}
\item \texttt{browser.click(element)} is a mouse shortcut for clicking an element
\item \texttt{browser.hover(element)} is a mouse shortcut for hovering a clickable element
\item \texttt{browser.drag(elementToDrag, elemenToDropOn)} is a mouse shortcut for dragging one element and dropping it to another
\item \texttt{browser.drag(elementToDrag, x, y)} is a mouse shortcut for dragging one element to absolutely specified coordinates
\item \texttt{browser.dragBy(elementToDrag, x, y)} is a mouse shortcut for dragging one element to relatively specified coordinates
\item \texttt{browser.choose(element, option)} is a mouse shortcut for selecting an option in a choosable form element
\item \texttt{browser.unchoose(element, option)} is a mouse shortcut for deselecting an option in a choosable form element
\item \texttt{browser.checkToggle(element, option)} is a mouse shortcut for toggling a checkable form element
\item \texttt{browser.chooseDate(element, date)} is a mouse shortcut for selecting a date in a datepicker form element
\item \texttt{browser.type(element, text)} is a keyboard shortcut for typing text into an input field
\item \texttt{browser.keyPress(element, key)} is a keyboard shortcut for pressing a single key while an element is focused
\item \texttt{browser.keyHold(key)} is a keyboard shortcut for holding a single key
\item \texttt{browser.keyRelease(key)} is a keyboard shortcut for releasing a single key
\end{itemize}

The \texttt{type(element, text)} method for example consists of two consecutive steps:
\begin{enumerate}
\item Querying for the typable elements and retrieving the first result: \texttt{browser.que\-ry(typable(element)).findFirst()}
\item If there is a result, typing the text in the typable element: \texttt{typableElement.ty\-pe(text)}
\end{enumerate}

The other shortcut methods perform in the same way.

\subsection{Imitating Human-like AJAX interactions}
\label{framework_javascript}
Many web pages use JavaScript (and so-called AJAX) to improve the user experience with rich user-interfaces.
AJAX applications move the UI logic to the client and some application logic as well (data manipulation and validation).
Web users often wait after clicking for a specific element to appear or disappear.
Abmash supports such interactions with the method \texttt{browser.waitFor().element(element)}, which searches every second for the target element. The execution continues when the element is found and contains the specified text or label.

For example, some web pages use JavaScript to show a login form without reloading the page.
Listing \ref{fd_wait} illustrates how one can interact with such login forms.
\begin{lstlisting}[caption={Wait for completion of asynchronous calls},label=fd_wait,float]
// initial click on the "login" text
browser.click("login");

// wait for any element labeled "Username"
browser.waitFor().element("username");

// more specific wait condition by using a query
browser.waitFor().element(browser.query(typable("username")));

\end{lstlisting}

If the query evaluates to false after a specifiable timeout, an exception is thrown. 
Furthermore, the framework automatically detects popup windows and alert dialogs. The currently active window can be selected by using \texttt{browser.window().switch\-To(windowName)} and  \texttt{browser.window().switchToMain()} to select the main window. Alerts with just one confirmation button are pressed immediately, whereas multiple buttons require the selection of one specific choice.

\subsection{Data Transformation}
\label{data_transformation}
Migration tasks often require transforming some pieces of data.
For instance, one may have to split a single address field into Street, City and Postcode.
Abmash does not contribute on this point (we refer the reader to \cite{DiLorenzo2009}) for discussion on this topic).
Since Abmash applications are embedded in a general purpose language, one can write all kinds of data transformation, the data transformation tasks can for instance be written in pure Java or using specific libraries (e.g., North Concepts' Data Pipeline).

\subsection{Prototype Implementation}
\label{application_architecture}

Let us now briefly present the prototype implementation of Abmash.

\subsubsection{Abmash and Selenium}
Abmash needs the visual rendering information of web pages, which is not available with libraries based on the document model such as HtmlUnit.
Since web pages are written, tested and optimized to be rendered on standard browsers, Abmash needs the same visual semantics as browsers. 
Hence, the visual rendering of web pages (size, position, color) is obtained directly from a browser (Mozilla Firefox in our case).
To get this information, we use Selenium.
Selenium is a browser automation library which aims at testing web applications from the end-user perspective.

\paragraph{How Selenium Works?}

Selenium is composed of two components which are independent from each other: a browser extension on top of standard browser extension mechanisms\footnote{e.g., for Mozilla Firefox extension, see \url{https://developer.mozilla.org/en-US/docs/Extensions}  \urlaccessed.}; and a Selenium server.
A Selenium server primarily runs test automation scripts, but also exposes a programming interface for third-party applications.
Abmash is a layer on top of this programming interface.

\paragraph{Which Part of Selenium Does Abmash Use?}
Abmash uses Selenium for two key mechanisms.
First, Abmash uses Selenium to trigger human like interactions such as clicking and keyboard input.
Second, Selenium is used to retrieve all the information to support the visual understanding of web pages. In particular, Selenium feeds Abmash with the exact coordinates in  pixel of every HTML element.

From a bird's eye view, Abmash programs are handled by three different components.
\emph{Abmash Core} handles browser control and command such as keyboard strokes and mouse movements as well as browser actions like handling multiple windows, popups and navigating through the browser history. 
The \emph{Visual Semantic Engine}  contains the logic to handle the queries based on the visual representation of the page (see Section \ref{visual_selection}). It interacts with the \emph{Selenium Binding} to find appropriate page elements to interact with.
The \emph{Visual Semantic Engine} is also responsible for computing the distance between page elements (see section \ref{framework_closeness}). The calculations are dependent on the given query predicates in order to provide the best result if a direction is specified.

\begin{figure}
\centering
\includegraphics[width=12cm]{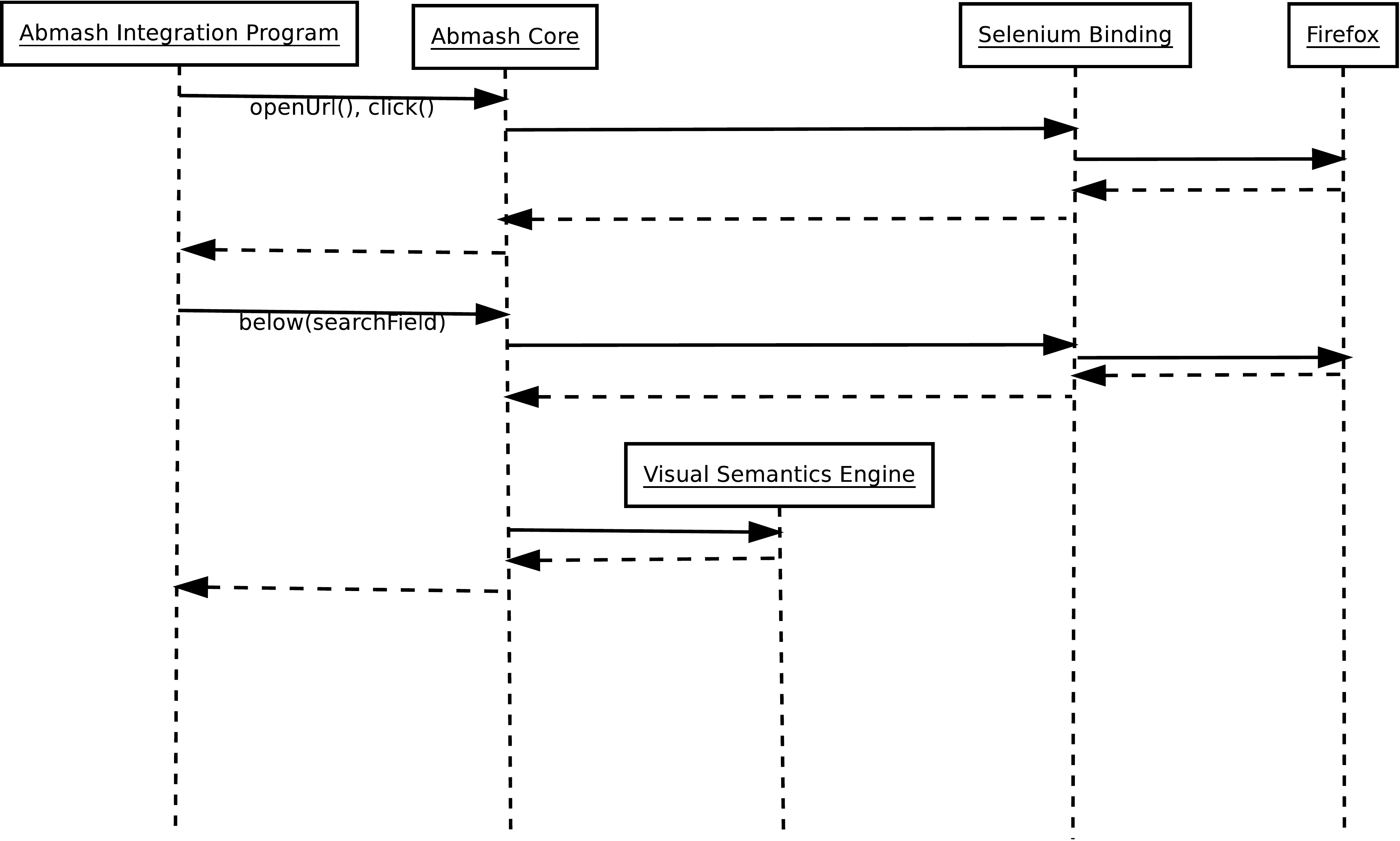}
\caption{Partial Sequence Diagram of the Abmash Program of Listing \ref{fd_first_example}}
\label{fig:seqdiag}
\end{figure}

The interactions between Abmash, Selenium, and the Firefox browser are illustrated by the sequence diagram of Figure \ref{fig:seqdiag}. When an Abmash integration program (such as those presented in this paper, e.g., \ref{fd_first_example}) calls \texttt{openUrl} or \texttt{click}, this command is translated into a call or a sequence of calls to the low level Selenium API, which are then routed to the browser (upper collaboration of Figure \ref{fig:seqdiag}).
When the developer uses queries based on the visual semantics (such as \texttt{below}), Abmash first retrieves all the required information from Firefox (coordinates, color, etc.) and the computes the results (set of HTML elements) with the \emph{Visual Semantic Engine}.
To sum up, Abmash provides a high-level, mashup-specific level of programming abstraction compared to the low-level routines of Selenium, and the even lower level routines of Firefox.

\subsubsection{How is the Visual Semantics Engine implemented?}
\label{visual_semantics_engine}

Let us now present how the visual semantics engine of Abmash is implemented.
The visual semantics engine is responsible for selecting the correct elements on the current web page according to visual semantic queries. After selection it's possible to interact with the returned elements.

\paragraph{Query Predicates}
\label{query_predicates}
Queries are formulated in Java (see section \ref{framework_element_type_predicates}). Queries consist of an arbitrary amount of predicates which are Java classes. Each predicate contains rules to create jQuery selectors and filters. When a query is executed, all predicates return their jQuery representation and the list of predicates is sent to the Abmash JavaScript library.

Internally, there are three categories of predicate evaluation:
First, the filter predicates are directly translated by simple jQuery selectors and filters. They mostly use the \texttt{find()} and \texttt{filter()} methods of jQuery.

Second, the visual predicates use the coordinates and dimensions of elements to compute direction or distance. In addition, elements that partially or completely cover the same space are detected as overlapping or hidden elements. Depending on the predicate parameters, the visual constraints on direction or distance need to match on at least one or all target elements (see also section \ref{framework_closeness}). Visual predicates consume more computation time and are therefore slower than filter predicates.

Third, color predicates analyze the visible colors of elements. A screenshot of the relevant part of the current page is made in Java and sent to the browser. The image is encoded as PNG Base64 string and loaded in an HTML \texttt{<canvas>} element. The image can then be analyzed pixel by pixel to check if the given elements match the dominance and tolerance parameters of the predicate (see also section \ref{color_predicate}). Color predicates are computation-intensive and also slower than filter predicates.

After creating all predicates, the generated selectors and visual filtering methods are executed. The resulting elements are collected, merged and ordered before the result is sent back to Java, where it can be used to extract information or to further interact with it within the mashup.

\paragraph{Merging Predicate Results}
Each predicate is evaluated in no particular order. The resulting elements are merged together depending on the given query. 
We have already presented how boolean predicates and nested predicates can be used to further filter a subset of predicate results.
Eventually, duplicates are always removed.

\paragraph{Distinct Descendants}
Due to the hierarchical nature of HTML, multiple elements can represent the same result. For example, if a paragraph \texttt{<p>} is wrapped by a block \texttt{<div>}, both elements are basically referring to the same semantic result. Abmash follows the principle of returning the most specific element in a hierarchy. If any element in the result set has a parent element with the same visible text that is also in the result set, the parent is removed from the result. This methodology prevents duplicates on a higher level and deals greatly with large result sets containing nested elements, especially for web pages with a large DOM tree.

\paragraph{Ordering Results}
The last step orders the elements by relevance. Depending on which query predicate has returned the element a weight is calculated (see also section \ref{ordering_result}). If multiple queries returned the element, the weight is further increased. Finally, the result is ordered by the descending weight.

\subsubsection{Running the Integration Logic}
The necessary infrastructure of Abmash is a Java Virtual Machine. 
The integration logic is run as a Java program. A main class takes the name of the Abmash class as parameter, together with configuration information if necessary, and runs the integration. This command line enables Abmash program to be automatically launched, for instance in a Crontab.

\subsubsection{Programming Language}
Abmash programs are intended to be written in Java, since the framework itself is written in Java.
We further discuss this point in section (\ref{discussion}).

\subsubsection{Other Dependencies}
In addition to depending on Selenium, it has the following optional dependencies: Abmash may use the Apache CXF library to automatically wrap human-like interactions as standard web services (this point is further discussed in \ref{sec:api-mashup}), and can use the Web User Interface library Vaadin to build improved web-based user-interfaces as an additional layer between existing web pages and Abmash applications.

\subsubsection{Extensibility}
Abmash is built to be extensible. For instance one can imagine adding a process engine in order to orchestrate human-like interactions together.
Abmash programs may be time-consuming hence applications can be parallelized to deal with integration mashups that are independent from each other. 
Note that parallelization requires more resources because multiple browser instances are launched at the same time\footnote{The system and hardware requirements of running a single Abmash runtime are a standard Java runtime environment and enough memory to execute the virtual machine in addition to the browser (typically 1 GB).}.

\subsubsection{Security Aspects of Abmash}
We now discuss the security aspects of Abmash. In short, since Abmash uses native browser windows, it has the very same security strengths and flaws as browsers.

Most browsers contain a security concept called ``Same Origin Policy'', which forbids scripts originating from external sources to manipulate the data of a page. This protects end users from attacks where malicious scripts are injected into the current website. 
Abmash does not break the policy since every page is opened in a native browser window. Abmash scripts are executed in the same context. That allows mashups for a maximum flexibility while profiting from all security measures of modern browser applications.

Another security issue with web applications is ``Cross-Site-Scripting'' (XSS). Attackers may inject malicious JavaScript code when user input is unchecked. Abmash is equally vulnerable to those types of attacks compared to normal usage of browsers. When opening pages in the Abmash browser, the mashup developer is responsible for checking the correctness of URLs and input data.
As Abmash uses native browser windows, it has the very same security strengths and flaws as browsers.
Abmash integration mashups in an EAI context are meant to be executed in an intranet setting, where proxies would protect the mashup from fetching or sending information from or to the Internet.

\section{Evaluation}
\label{evaluation}

This section presents an evaluation of the Abmash integration framework for web applications.
The evaluation is composed of:
first, a thorough presentation of proof of concept applications built with Abmash (Sec. \ref{case_study});
second, the result of a user study conducted to assess the usability of the framework (Sec. \ref{user_experiment});
third, a comparative study of Abmash code against another technique for imitating human interactions  (Sec. \ref{comparative_study});
finally, an evaluation of the performance of the framework in terms of execution time (Sec. \ref{performance}).
We also discuss the limitations of the approach in Section \ref{discussion}.

\subsection{Proof of Concept Applications}  
\label{case_study}

The following presents proof-of-concept Abmash applications tackling realistic concrete problems.
These example applications illustrate some areas of application of the Abmash framework: content migration, and API creation.
They aim at showing the capabilities of the framework.

\subsubsection{Content Migration from Legacy Web Applications}
\label{case_study_migration_mashup}

In the IT landscape of companies, certain applications contain important and valuable data that need to be persisted when migrating to a different (generally newer) system.
Often, the legacy (source) application that is put out of service has no API to export the existing data. Similarly, it happens that the destination application has no programmatic import API.
For example, the whole scenario is daily experienced by companies migrating from one Content Management System (CMS) to another.
The following presents how the Abmash framework can be used to migrate the content of one blog platform to another one in an automated and effective manner.

\paragraph{Scenario}
The scenario consists of transferring the content from one blog application to another. The source blog engine is SweetRice\footnote{SweetRice: \url{http://www.basic-cms.org/} \urlaccessed} while the target engine is Croogo\footnote{Croogo: \url{http://croogo.org/} \urlaccessed}. The SweetRice blog is initially filled with fake articles and comments and Croogo does not have any content. The scenario is realistic, because it is common to replace applications in order to match new business requirements.
It is to be noted that both blog engines are unmodified and do not provide a comprehensive programming interface for exporting or importing data.

To complete the migration, the following content has to be moved from the source to the target blogs.
The articles: each article has a title and a text, they are written by an author.
The comments: articles may have comments, each comment also has an author.

\paragraph{Approach}
The migration of one article with the associated data consists of two successive steps: data extraction and data submission. For the extraction, element selection needs to be done (see section \ref{framework_description}). Data submission is achieved by logging in as an administrative user, navigating to the submission form and filling in the previously extracted data. Browser instances for each blog are used to imitate the users behavior of using several windows or open tabs to achieve the task.

\begin{lstlisting}[caption={Find blog articles},label=lst:sweetrice_findarticles, float]
// navigate to Sitemap
browserSource.click("Sitemap");

// find all article titles by identifying clickable elements
// above the footer
HtmlElement footer = browserSource.query(contains("copyright")).findFirst();
HtmlElements articleTitleElements = browserSource.query(clickable(), leftOf(image()), above(contains("copyright"))).find();(*@\label{code:sweetrice_findarticles}@*)
\end{lstlisting}

\begin{lstlisting}[caption={Extract article text and date},label=lst:sweetrice_extract, float]
HtmlElement textElement = browserSource.query(below(titleElement), text()).findFirst();
HtmlElement infoElement = browserSource.query(below(textElement), text()).findFirst();
String text = textElement.getText(); 
Date date = infoElement.extractDate();
\end{lstlisting}

\begin{figure}
\centering
\includegraphics[width=10cm, viewport=0 0 510 161]{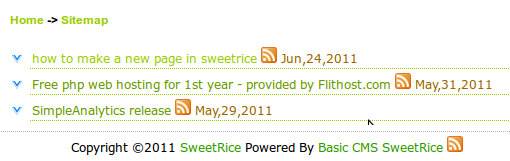}
\caption{Sitemap of SweetRice which contains a list of all articles}
\label{fig:sweetrice_sitemap}
\end{figure}

SweetRice per default shows the latest three articles on its main page. A complete list of all articles can be seen on the Sitemap of the blog. Each element in the list consists of a link to the article, a feed icon and the publishing date (see figure \ref{fig:sweetrice_sitemap}).
With Abmash (see listing \ref{lst:sweetrice_findarticles}), the article links can be found by searching for all clickable elements which are located left to the feed icons. To avoid selecting the links ``SweetRice'' and ``Basic CMS SweetRice'' in the footer, a predicate to ignore all elements in the footer is added (see listing \ref{lst:sweetrice_findarticles}, line \ref{code:sweetrice_findarticles}):
Then, the text of the blog post and creation date are fetched by queries based on visual closeness (\emph{below} followed by \emph{findFirst}), as shown in listing \ref{lst:sweetrice_extract}. 
The comments are located on another page and can easily be identified through the proceeding number sign ``\#'' (see 
listing \ref{lst:sweetrice_comments}).
\begin{lstlisting}[caption={Find article comments},label=lst:sweetrice_comments,float]
// comments in sweetrice are labeled #1, #2, etc
HtmlElements referenceElements = browserSource.query(clickable("#")).find();
for (HtmlElement referenceElement: referenceElements) {
  HtmlElement commentInfoElement = browserSource.query(rightOf(referenceElement), text()).findFirst();
  HtmlElement commentTextElement = browserSource.query(rightOf(commentInfoElement), text()).findFirst();

  String commentAuthor = commentInfoElement.getTextFirstLine();
  Date commentDate = commentInfoElement.extractDate();
  String commentText = commentTextElement.getText();
}
\end{lstlisting}

\begin{figure}
\centering
\includegraphics[width=10cm, viewport=0 0 659 141]{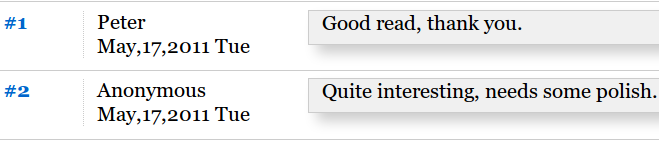}
\caption{Article comments in SweetRice}
\label{fig:sweetrice_comments}
\end{figure}

\begin{lstlisting}[caption={Login to admin interface},label=lst:croogo_login,float]
browserTarget.type("Username", "admin");
browserTarget.type("Password", "Ab!2cDe").submit();
\end{lstlisting}

Croogo is secured with a login system to prevent unauthorized persons from adding or removing content. Since we have the login username and password, it is possible to automatically login by using keyboard interactions on the form elements (see listing \ref{lst:croogo_login}).
The creation of a new article is achieved by a sequence of navigation and input interactions. First, one navigates to the form which allows us to add a new article. 
Then, the title and the text are entered in text input fields, whereas the creation date is chosen from dropdown boxes and located in the ``Publishing'' tab.
The whole process is shown in listing \ref{lst:croogo_enter}.
Comments are eventually added by filling in the comment form in the frontend of Croogo as normal human users would do (see listing \ref{lst:croogo_comment}).

The entire application is written in just over 50 lines of code and is capable of migrating the content from the SweetRice to the Croogo blog engine. The resulting code reflects all interactions a human user would do to migrate the data, without any knowledge of the source code of both web applications.
Not all the data can be migrated by imitating human interactions.
For instance, in this proof of concept implementation of a blog migration, we were not able to migrate the commenter emails and the comment dates.
The former is due to the fact that this information, while in the database, is not present in the SweetRice interface to prevent spiders to collect mail addresses for spam purposes.
The latter is due to the fact that there is simply no field to set the comment date.
This shows that  migration by imitating human interactions is no silver bullet. However, most of the migration is still automated without interacting with the source and target applications at the source or database level.

\begin{lstlisting}[caption={Enter and submit article data},label=lst:croogo_enter,float]
// save blog article at new blog page
// being logged in as administrator, navigate to creation of new blog article
browserTarget.click("Content");
browserTarget.click("Create content");
browserTarget.click("Blog");

// article content
browserTarget.type("Title", title);
browserTarget.type("Body", text);

// set publishing date
browserTarget.click("Publishing");
browserTarget.chooseDate("created", date);

// save blog article
browserTarget.click("Save");
\end{lstlisting}

\begin{lstlisting}[caption={Enter and submit comment data},label=lst:croogo_comment,float]
browserTarget.click(title);
browserTarget.type("Name", commentAuthor);
browserTarget.type("Email", commentAuthorEmail);
browserTarget.type("Body", commentText).submit();
\end{lstlisting}

\subsubsection{API Mashup}
\label{sec:api-mashup}
Popular web applications often offer APIs to use their services programmatically. For example, Google offers an API to read and write appointments from Google Calendar and Facebook allows remote access to the wall of an authorized API user. Nevertheless, many legacy web applications do not offer such interfaces,.

For integrating a legacy web application with other applications, a possibility is to first wrap the legacy application into an API, and then design and implement the integration using this new wrapper API.
For instance, one can wrap a legacy CRM into a SOAP interface to add, modify and delete customers from the customer database.
We use the term \emph{API Mashup} to refer to such APIs, created on top of legacy web applications.

\paragraph{Scenario}
Trac\footnote{Trac: \url{http://trac.edgewall.org/} \urlaccessed} is an issue tracker for software development teams. It offers no API, hence difficulties may arise if developers want to check or update issues remotely.
We would like to build RESTful\footnote{REST Web Service} services for different tasks related to issue tracking:
\begin{itemize}
\item Fetch latest tickets;
\item Enter new tickets;
\item Search for tickets;
\item Get and create milestones;
\item Read and modify documentation pages.
\end{itemize}

\begin{lstlisting}[caption={Getting tickets from Trac with visual queries},label=lst:trac_api_get_tickets, float]
public class Ticket { 
  // data class modeling a ticket
  // with getters and setters
  ...
}

public class TracAPI {
  public Tickets getTickets(int count) {
    // navigate to the list of latest tickets
    browser.openUrl("http://trac.myproject.com/trac/report/1?max=" + count);

    Tickets tickets = new Tickets();
    
    // find and store the ticket table
    Table ticketTable = browser.getTable("Ticket");

    // iterate through each table row and save the data
    for (TableRow row: ticketTable) {
      Ticket ticket = new Ticket();
      // all getText calls select the corresponding column
      // of the row (trac being set up in the German language)
      ticket.setId(row.getText("Ticket"));
      ticket.setSummary(row.getText("Kurzbeschreibung"));
      ticket.setComponent(row.getText("Komponente"));
      ticket.setVersion(row.getText("Version"));
      ticket.setMilestone(row.getText("Meilenstein"));
      ticket.setType(row.getText("Typ"));
      ticket.setOwner(row.getText("Verantwortlicher"));
      ticket.setStatus(row.getText("Status"));
      ticket.setCreated(row.getText("Created"));
      tickets.add(ticket);
    }

    return tickets;
  }
}
\end{lstlisting}

\paragraph{Approach}
Exposing the tickets of Trac as a web service requires to first extract the tickets from the web interface and then to wrap them as a RESTful web service.
The extraction is done by using the framework routines related to table extraction. This is shown in listing \ref{lst:trac_api_get_tickets}.
One first retrieves the table containing the tickets (the table is preceded by a label ``Tickets''. Then, for each row of the table, one gets the text in specific columns. For instance, \verb|row.getText("Kurzbeschreibung")| retrieves the summary (Kurzbeschreibung in German) of the ticket as the cell in column ``Kurzbeschreibung''.
The tickets are then exposed as a RESTful web service using the  Apache CXF library. Apache CXF is based on annotations to specify what and how to expose.
Listing \ref{lst:trac_api_webservice} shows that the ticket web service can be achieved with less than 30 LOC.

\paragraph{Comparison with Another Unofficial API}
\label{comparative_study_tracapi}
\textit{Trac} has no official API. There is an unofficial, not actively maintained XML-RPC plugin for Trac\footnote{Trac XML-RPC Plugin: \url{http://trac-hacks.org/wiki/XmlRpcPlugin} \urlaccessed}. 

The functionalities of the prototype API Mashup and this other API are very similar. For instance, both allow the user to fetch a filtered set of currently available tickets and to enter new tickets. The plugin is also able to remotely read and modify wiki pages, whereas the Mashup offers a search API. 

The development of the Trac XML-RPC plugin required extended knowledge about the implementation details of the Trac application which is written in the Python programming language. Plugins extend the built-in functionality by creating implementations of already existing extension interfaces. 
The plugin developer had to know about these components, extension points and possibly the database structure before being able to develop the plugin. 
In contrast, the Abmash API Mashup presented in this section is able to provide an API, with no knowledge of the plugin architecture, extension points, source code and database schema of Trac. All functionalities are implemented as an interaction with the presentation UI layer of Trac. Thus, the Abmash API is very similar to the functionality of the Trac plugin while being much easier to develop.

\begin{lstlisting}[caption={Creating an API Mashup with Apache CXF Annotations},label=lst:trac_api_webservice,float]
@WebService(endpointInterface = "com.abmash.webservice.trac.TracService")
@Produces(MediaType.APPLICATION_XML)
@Consumes(MediaType.APPLICATION_XML)
@Path("/tracService/")
public class TracService {
  Browser browser;

  @GET
  @Path("/tickets/{count}")
  public Tickets getTickets(@PathParam("count")int count) {
    [...] // 10 lines of glue code
    TracAPI tracAPI = new TracAPI(browser);
    return tracAPI.getTickets(count);
  }
\end{lstlisting}

\subsection{User Study}
\label{user_experiment}

This section presents the design and the results of a user study that we conducted to assess the usability of the Abmash framework.
We want to make sure that developers can intuitively understand the framework's goal and usage.
With this study, we aim at answering the following questions:
\begin{itemize}
\item Can developers intuitively understand how to use the framework?
\item Are they able to create a meaningful application which delivers correct results?
\item Is there some missing  functionality which hinders or slows down the development of the integration program?
\end{itemize}

\subsubsection{User Study Design}

The user is asked to implement a migration mashup consisting of migrating the content from one blog engine to another one.
In other terms, the user is asked to write a mashup that is similar to the prototype introduced in section \ref{case_study_migration_mashup}, yet greatly simplified.
Indeed, the user is asked to only migrate the post title and content, and not the authorship information, the comments, etc.
The user is given 45 minutes to solve the task and write an application that successfully migrates the required content.
While the task is feasible in 45 minutes, it still resembles a realistic scenario.

The user is allowed to only use the Abmash sources, its Javadoc, the default JDK packages and a preconfigured Eclipse environment.
He is asked to solve the task while examining the HTML source code as little as possible.
At the end, he is asked for feedback through a live discussion and a questionnaire.
The collected information is important to evaluate the framework usefulness regarding different development approaches and styles to solve this particular problem.
As in section \ref{case_study_migration_mashup}, the source blog engine is SweetRice 0.7.0 and the target blog engine is Crogoo 1.3.3.

The study is controlled is the sense that:
\begin{enumerate}
\item the user has to solve the tasks in a fixed order;
\item he is given a fixed development environment;
\item he is  being observed by the experimenter during the task.
\end{enumerate}

The replication material is publicly available at \url{http://goo.gl/tIXyX}.

\subsubsection{Participants}

Five subjects participated to the user study.
They were graduate students in computer science at the Technische Universität Darmstadt.
They had no previous knowledge of the Abmash framework. 
They have between 2 and 8 years of experience in Java development as freelancer or part-time developers.
Their experience with web development and especially HTML and CSS markup varies strongly: two participants state that they have nearly no knowledge about web development whereas the others estimate their skills as above-average. 

\subsubsection{Results}
Four out of five participants solved the study's task and developed a working migration application within 45 minutes. 
In a nutshell, they declared having found the framework intuitive and having had fun while programming by imitating human actions.
Our live observations confirm these statements.
More precisely, we make the following observations.

\paragraph{Motivation}
The participants were motivated to complete their assignment.
As reported by them, the reasons are that
\begin{enumerate}
\item they made rapid progress in understanding the framework's approach
\item they can watch the execution of the browser automation commands, it gives an immediate visual feedback about the mashup.
\end{enumerate}

\paragraph{Imitation of human actions}
All developers initially looked at the HTML source code, but most understood the ``human-like'' concept after some trials-and-errors, and they quickly adopted the Abmash style to find and interact with page elements.
Four subjects correctly used the visual query methods to find elements close to another in a specific direction, but needed some time to realize that they exist and how they work.
The one participant who did not understand this key part of the framework had advanced knowledge on web development and used the HTML source code intensely.
He really wanted to stick to the document HTML structure (the DOM) and did not understand why the corresponding low level DOM manipulation methods do not exist at all in Abmash.
Interestingly, the participants which felt most comfortable with the framework's concept were those with no or little experience in HTML/CSS.
According to our observations, we think that the more web development experience a programmer has, the more time and effort are necessary to understand the differing approach of the framework. 
We also think that this emphasizes the ``mashup taste'' of Abmash: users with no knowledge in web development can still work with and integrate legacy web applications.

\paragraph{Documentation}
The biggest issues in achieving the task were caused by incomplete descriptions in the documentation. 
In particular, criticism was given against non-descriptive Java exceptions.
Also, the participants generally missed examples and code snippets. 

\paragraph{Problem Solving}
The participants were free to decide how to create the mashup.
Indeed, they had very different approaches to solve the task. Most of them used a step-by-step bottom-up strategy (running small pieces of code), but there was one participant who created the application in a top-down manner (writing the mashup as whole before debugging). Both attempts worked well. Indeed, the Abmash framework is meant to enable developers to express their creativity and not to constrict their problem solving approaches.

\paragraph{Framework Experience}
As part of the study, all participants had to answer questions about their feelings with the framework after the development phase. They were asked to give certain aspects of the framework a score on a discrete scale, which contained the answers ``yes, definitely (10)'', ``partially (7)'', ``just barely (4)'' and ``not at all (1)'' (see Table \ref{tab:questionnaire}). 
In general, the Abmash framework was considered as well designed. The resulting Java code was considered readable, and understandable. The participants intuitively understood the fluent interface to chain commands and used them frequently. Especially the chaining of visual predicates to find the desired web page elements was considered as a positive experience. Overall, the framework was considered as encouraging rapid prototyping.
Also, the goal of the framework (human imitation) was considered interesting and the participants felt that the chosen scenario for the study is very suitable. 
One participant said that one hour is too short to make a valid statement about the quality of the framework design.

\begin{table}
\begin{tabularx}{\textwidth}{|X|p{2cm}|}
\hline
\textbf{Question} & \textbf{Score}\\
Is the framework suitable for the task? & 9\\
Is the framework easy to understand? & 8\\
Is the framework easy to use? & 9\\
Is the framework intuitive? & 8,5\\
Does it provide the expected results? & 8,5\\
Is the documentation sufficient? & 7\\
Is the framework well-designed? & 10\\
Do you find the goal of the framework interesting? & 9,5
\\\hline
\end{tabularx}
\caption{Average Results of the Questions Given at the End of the User Study}
\label{tab:questionnaire}
\end{table}

\paragraph{Suggestions}

The participants had raised many interesting ideas about the framework.
\begin{itemize}
\item The most requested feature was visual predicates based on colors (e.g. \texttt{isRed()}), which would further support the imitation of human interactions.

\item Many web applications are sensitive to the language configuration of the browser.
It has been suggested being able to set a language, in order to avoid spurious breaking of the automated mashup.

\item It also has to be considered that all visual methods depend on the current browser window size. Many web pages have a dynamic design that uses the space available to display the content. Thus, a smaller viewport would lead to rearranged elements and different results for visual queries. It has been suggested to force setting the browser window size.

\item Users asked for improving the console output, because it could not be clearly determined whether the mashup is still running or not. This led to some confusion during the user study.

\item Since the framework goal is to write mashups, some participants suggested building a graphical user interface and a graphical language to allow non-programmers to develop mashups (i.e. providing some kinds of end user programming support).

\end{itemize}

Since that experiment, most limitations have been overcome: the documentation has greatly been improved and some suggestions such as color queries have been implemented. 

\subsubsection{Threats to Validity}

This user-study was a first assessment of the usability of the Abmash framework.
Four out of 5 subjects successfully performed the task in less than 45 min and the overall feedback was very positive.
Let us now discuss the threats to validity.

\paragraph{Threats to Construct Validity}

Our experiment may not only measure the ease of use of the Abmash framework.
One key factor may interfere with the results:
the participants were enthusiastic to participate to a user experiment on a voluntary basis. Hence, their learning motivation may have been much higher than the one in a company setup.
Also, since we both designed Abmash and the task of the user study, there may be a bias towards a too appropriate task.
We were careful in minimizing this threat by setting up a realistic task (blog migration) involving off-the-shelf unmodified web applications.

\paragraph{Threats to External Validity}
While the results are quite clear, they may not be completely generalizable. First, the experiment involved a small number of participants.
Second, they all share the same background (graduate students in computer science at the same university).
We are aware of this risk but since our open-source prototype has received some attention, we are confident in the generalizability of the ease-of-use of Abmash.

\subsection{Comparative Study}
\label{comparative_study}

This section provides comparisons between Abmash and other approaches on concrete issues that often arise with mashup development.

\subsubsection{On Dynamically Generated Forms}

\begin{lstlisting}[language=html,caption={The issue of generated DOM attributes. One can not write CSS/Xpath selectors based on generated attributes.},label=listinggenerated,float] 
<form name="book" action="process" method="post">
  <div>
    <label for="autogenerated101">Author</label>
    <input id="autogenerated101" name="autogenerated101" type="text" />
  </div>
  <div>
    <label for="autogenerated201">Title</label>
    <input id="autogenerated201" name="autogenerated201" type="text" />
  </div>
  <div>
    <input id="autogenerated301" name="autogenerated301" type="submit" value="Save" />
  </div>
</form>
\end{lstlisting}

Most web applications generate HTML content on the fly.
Furthermore, many of them use dynamically generated identifiers, as shown in Listing \ref{listinggenerated}.
In such cases, it is hard to interact with the application at the level of the HTML structure because the identifiers are not a priori known.

In the presence of dynamically generated identifiers, Abmash proves to be very robust.
The Abmash selector strategies are able to identify all input elements of a page by their corresponding labels. 
The developers have an easy access to HTML elements without knowing their exact internal id or location in the document structure.
Listing \ref{fd_forms} shows that  with Abmash, the content can be read and submitted without any knowledge of the HTML source code.
Indeed, the mashup is not sensitive when the underlying HTML structure changes and the Abmash query methodology is more robust against changes in the document structure compared to static query on the HTML DOM structure with Xpath or CSS queries.
if there would have been multiple input fields with the same ``Author'' label, the programmer would have to use another visual specifier such as \texttt{first} or 
\texttt{below} to uniquely identify the field.

\begin{lstlisting}[caption={Handling dynamically generated forms},label=fd_forms,float]
// type in author and title in the corresponding input fields
// just by addressing their labels, and finally submit the form
browser.type("Author", "J.R.R. Tolkien");
browser.type("Title", "Lord of the Rings").submit();
\end{lstlisting}

\subsubsection{On Directly Writing Selenium Code}

Abmash is built on the Selenium 2 framework. In this section, we explore whether using the Abmash framework adds any value compared to directly using Selenium.

\paragraph{Identifying Specific Elements by Visible Element Properties}

Finding web page elements is one of the most important tasks when interacting with web pages.
For instance, finding the products of an e-commerce website means finding elements that have a price.
Listing \ref{listing_query_abmash} shows a default query a mashup developer can start with.
With Abmash, querying for web page elements can be done by chaining an arbitrary number of search criteria. Each predicate aims to find elements by means of visible attributes to imitate human behavior when looking for the correct elements to interact with. 

The  corresponding Java code using the Selenium API is much longer\footnote{An excerpt is available at \url{https://gist.github.com/3761416} \urlaccessed}.
In Selenium, every possible selector needs to be executed separately. Furthermore, the order of execution is important and can lead to wrongly selected elements.

\begin{lstlisting}[caption={Element queries with \textbf{Abmash}},label=listing_query_abmash, float]
Browser browser = new Browser("http://example.com/some/page");
HtmlElements clickableElements = browser.query(clickable("price")).find();
\end{lstlisting}

\paragraph{Tables and Lists}

Tables and lists are common representations in web applications. Extracting specific parts of their content is usually difficult without knowledge of the document structure.
Selenium does not detect table structures by default and therefore needs the development of table handling code\footnote{An excerpt is available at \url{https://gist.github.com/3761429} \urlaccessed}. 
On the contrary, as shown in Listing \ref{listing_abmash_table}, Abmash can find table structures by searching for a keyword in the visible content of any table and returns the table representation.

\begin{lstlisting}[caption={Table data extraction with \textbf{Abmash}},label=listing_abmash_table,float]
// Find desired table and extract the data from the specified cell
Browser browser = new Browser("http://example.com/some/page");
Table userTable = browser.getTable("Username");
HtmlElement cellInSecondRowAndMailColumn = userTable.getCell(1, "Email");
\end{lstlisting}

\paragraph{Boolean Selector Queries}

Complex element queries may be indispensable to interact with the right set of web page elements. Combining multiple subqueries would give the developer more control over the browser interactions.
Selenium does not allow the user to query for elements at a fine-grained level of detail, except for concrete CSS and XPath selectors. These selectors are no option if there is no knowledge about the document source code, so finding the elements has to be coded by hand. In addition, the boolean expressions need to be composed manually\footnote{An excerpt is available at \url{https://gist.github.com/3761440} \urlaccessed}. 

On the contrary, Abmash offers a simple interface to create complex queries while keeping the human imitation approach (see Listing \ref{listing_complex_queries_abmash}). The queries are readable and rather intuitive. This provides an advantage for development efficiency and maintainability.

\begin{lstlisting}[caption={Complex queries with \textbf{Abmash}},label=listing_complex_queries_abmash,float]
Browser browser = new Browser("http://example.com/some/page");

HtmlElements elements = b.query(
  clickable(),
  below(contains("Information")),
  or(
    image(),
    rightOf("Language"),
    not(contains("english"))
  )
).find();
\end{lstlisting}

\paragraph{Extraction of Date/Time Information in Various Formats}

Information on the web is moving very fast and is often connected to timestamps. Articles have a creation date, events have start and end dates and micro-blogging status updates can be tracked in real-time. 
Selenium does not offer any options to extract temporal contents out of web pages, which leaves it to the developer to write the code completely by himself\footnote{An excerpt is available at \url{https://gist.github.com/3761448} \urlaccessed}. 
Abmash provides developers with solutions which cover most standards of date/time formats.
Listing \ref{listing_abmash_date} shows a default snippet that extracts the date out of a textual element which contains the label ``Created''.

\begin{lstlisting}[caption={Extraction of dates with \textbf{Abmash}},label=listing_abmash_date,float]
Browser browser = new Browser("http://example.com/some/page");
HtmlElement elementWithDate = browser.query(text("Created")).findFirst();
Date date = elementWithDate.extractDate();
\end{lstlisting}

\paragraph{Summary}

Even if Selenium is technically capable of most use cases that can be done with Abmash, this comparative study has shown that pure low-level Selenium code can be rather verbose and complicated.
For mashups based on imitating human interactions, Abmash code is generally much easier to read, write and maintain compared to raw Selenium code.

\subsection{Performance Evaluation}
\label{performance}

We now focus on the evaluation of the performance of Abmash migration mashups.
The experimental methodology is as followed:
first, we list likely queries that span all features of Abmash,
second, we select real-world web pages,
third, for each pair of query and web page, we measure the execution time of Abmash.
We concentrate on using complex queries on real web pages in order to have a realistic estimation of Abmash's performance.

\subsubsection{Queries}
The queries used in the experiment are as follows (they are all explained in Section \ref{visual_selection}):
basic queries (\texttt{headline()}, \texttt{clickable()}, \texttt{typable()} or \texttt{image()});
queries with lookup of values  (e.g. \texttt{headline("jquery")});
color-based queries (\texttt{color(ColorName.WHITE)});
visual direction based queries (\texttt{below(headline("jquery"))});
complex boolean queries that combine simple, color and direction predicates.
The 20 queries are listed in Table \ref{table:performance_query}.
Those queries span the features of Abmash and are representative of client usages. 

\subsubsection{Web Pages}
We select 5 web pages for executing the queries aforementioned.
The inclusion criteria are:
first, they come from real world applications of the Web;
second, they are complex in the sense that they either have a large number of DOM elements or make heavy use of JavaScript and DOM manipulation.

The web pages are:
\begin{enumerate}
\item The Google Search results page for "jquery"
\item The Official jQuery documentation on filters \footnote{\url{http://api.jquery.com/filter/}}
\item A blog article about the performance of jQuery \footnote{\url{http://www.jquery4u.com/jsperf/jquery-performance-dom-caching/}}
\item The Wikipedia page about jQuery\footnote{\url{http://en.wikipedia.org/wiki/Jquery}}
\item The Stackoverflow page that lists questions tagged with "jquery"\footnote{\url{http://stackoverflow.com/questions/tagged/jquery}}. Stackoverflow  is a popular question and answer (Q\&A) site for programmers.
\end{enumerate}

As descriptive statistics, the first line of Table \ref{table:performance_query} gives the number of DOM elements of those pages, it ranges from 517 to 1582. 

\begin{table}

 \begin{center}
\begin{tabular}{l|p{4cm}|r|r|r|r|r||r}
& 																											& {Google} 				& {jQ Docs} 				& {c|}{Blog} 				& {Wikip.} 				& {Stackov.} 			& Avg. \\  \hline 
&{\scriptsize DOM element count in total} 																	& {{\scriptsize 517}} 	& {{\scriptsize 889}} 	&{{\scriptsize 683}} 	& {{\scriptsize 1582}} 	& {{\scriptsize 763}} 	& {\scriptsize 887} \\  \hline\hline 
1.& {\scriptsize \texttt{headline()}}                                    &  \textbf{1}s     &  \textbf{795}ms  &  \textbf{321}ms  &  \textbf{560}ms    &  \textbf{1}s     &  <\textbf{1}s \\ \hline 
2.& {\scriptsize \texttt{clickable()}}                                     &  \textbf{3}s     &  \textbf{1}s     &  \textbf{1}s     &  \textbf{10}s    &  \textbf{3}s     &  \textbf{4}s \\ \hline 
3.& {\scriptsize \texttt{typable()}}                                     &  \textbf{211}ms  &  \textbf{120}ms  &  \textbf{118}ms  &  \textbf{114}ms    &  \textbf{80}ms   &  <\textbf{1}s \\ \hline 
4.& {\scriptsize \texttt{checkable()}}                                     &  \textbf{92}ms   &  \textbf{108}ms  &  \textbf{90}ms   &  \textbf{115}ms    &  \textbf{94}ms   &  <\textbf{1}s \\ \hline 
5.& {\scriptsize \texttt{choosable()}}                                     &  \textbf{71}ms   &  \textbf{120}ms  &  \textbf{113}ms  &  \textbf{115}ms    &  \textbf{83}ms   &  <\textbf{1}s \\ \hline 
6.& {\scriptsize \texttt{submittable()}}                                   &  \textbf{527}ms  &  \textbf{120}ms  &  \textbf{160}ms  &  \textbf{130}ms    &  \textbf{154}ms  &  <\textbf{1}s \\ \hline 
7.& {\scriptsize \texttt{image()}}                                       &  \textbf{125}ms  &  \textbf{122}ms  &  \textbf{138}ms  &  \textbf{132}ms    &  \textbf{180}ms  &  <\textbf{1}s \\ \hline 
8.& {\scriptsize \texttt{contains("jquery")}}                                &  \textbf{6}s     &  \textbf{5}s     &  \textbf{1}s     &  \textbf{6}s     &  \textbf{3}s     &  \textbf{4}s \\ \hline 
9.& {\scriptsize \texttt{headline("jquery")}}                                &  \textbf{2}s     &  \textbf{1}s     &  \textbf{519}ms  &  \textbf{1}s     &  \textbf{774}ms  &  \textbf{1}s \\ \hline 
10.& {\scriptsize \texttt{clickable("jquery")}}                                &  \textbf{26}s    &  \textbf{5}s     &  \textbf{3}s     &  \textbf{5}s     &  \textbf{4}s     &  \textbf{9}s \\ \hline 
11.& {\scriptsize \texttt{image("jquery")}}                                  &  \textbf{3}s     &  \textbf{4}s     &  \textbf{7}s     &  \textbf{7}s     &  \textbf{14}s    &  \textbf{7}s \\ \hline 
12.& {\scriptsize \texttt{color(ColorName.WHITE)}}                               &  \textbf{18}s    &  \textbf{65}s    &  \textbf{19}s    &  \textbf{46}s    &  \textbf{29}s    &  \textbf{35}s \\ \hline 
13.& {\scriptsize \texttt{color(ColorName.BLUE)}}                              &  \textbf{9}s     &  \textbf{37}s    &  \textbf{15}s    &  \textbf{23}s    &  \textbf{13}s    &  \textbf{19}s \\ \hline 
14.& {\scriptsize \texttt{contains("jquery"). clickable()}}                          &  \textbf{3}s     &  \textbf{3}s     &  \textbf{715}ms  &  \textbf{3}s     &  \textbf{2}s     &  \textbf{2}s \\ \hline 
15.& {\scriptsize \texttt{clickable("jquery"). image()}}                           &  \textbf{25}s    &  \textbf{4}s     &  \textbf{3}s     &  \textbf{4}s     &  \textbf{4}s     &  \textbf{8}s \\ \hline 
16.& {\scriptsize \texttt{clickable("jquery"). color(ColorName.WHITE)}}                    &  \textbf{30}s    &  \textbf{15}s    &  \textbf{6}s     &  \textbf{13}s    &  \textbf{9}s     &  \textbf{15}s \\ \hline 
17.& {\scriptsize \texttt{below(headline("jquery")). clickable()}}                       &  \textbf{140}s   &  \textbf{74}s    &  \textbf{10}s    &  \textbf{73}s    &  \textbf{170}s   &  \textbf{93}s \\ \hline 
18.& {\scriptsize \texttt{below(typable()). above(clickable("jquery")). headline()}}             &  \textbf{241}s   &  \textbf{59}s    &  \textbf{24}s    &  \textbf{4}s     &  \textbf{243}s   &  \textbf{114}s \\ \hline 
19.& {\scriptsize \texttt{below(headline("jquery")). clickable(). color\newline (ColorName.WHITE)}}      &  \textbf{139}s   &  \textbf{73}s    &  \textbf{13}s    &  \textbf{83}s    &  \textbf{166}s   &  \textbf{95}s \\ \hline 
20.& {\scriptsize \texttt{below(headline("jquery")). clickable(). not(color\newline (ColorName.WHITE))}}   &  \textbf{132}s   &  \textbf{96}s    &  \textbf{25}s    &  \textbf{99}s    &  \textbf{170}s   &  \textbf{104}s \\ 
\end{tabular}
\caption{Query performance on different web pages. The numbers in bold is the time (in seconds of milliseconds) for executing a query (row) on a web page (column).}
\label{table:performance_query}
 \end{center}
\end{table}

\subsubsection{Experimental Results}
Table \ref{table:performance_query} gives the results of this experiment.
The main columns correspond to the 5 web pages.
There is one line per query.
A cell in the table is the execution time of a given query on a given web page.

Simple queries like \texttt{headline()} (queries \#1-\#7) without lookup strings mostly execute in less than a second. 
An exception is the query ``clickable()'' on wikipedia which lasts 10 seconds. This is due to the fact that the clickable predicate also searches for all elements that have an ``onclick'' attribute, which takes more time because all elements on the page need to be iterated through. The wikipedia page has the most DOM elements and links which makes the query slower than on the other test pages.
Looking-up strings (queries \#8-\#11) like in \texttt{contains("jquery")} takes between 4s and 9s. 
This is because analyzing the inner text of elements takes additional computation time. 
The query \texttt{clickable("jquery")} on the Google Search page performed significantly slower (26 seconds) than on the other pages. This is because searching for clickable elements with a lookup string involves implicit direction matching, for example to search for clickable checkboxes with a descriptive label close to it. On the Google page, there are more input elements than on the other test pages. Direction matching is slower than simple predicates, which explains the performance differences.

Color queries (queries \#12-\#13)  takes 19s and 35s. They are slower than simple queries because they require a screenshot, which gets loaded in the browser and analyzed in the query execution process. 
Queries \#14-\#16 combine the basic query types. Their execution time is between 2s and 15s.
Note that the composite query \texttt{clickable("jquery"). color(ColorName.WHITE)} runs faster than \texttt{color(ColorName.WHITE)} because the first query enables the engine to only analyze the colors of a subset of elements. 
This means that it is good to write more specific query predicates first to obtain faster query executions.

Queries \#17-\#20 involve direction analysis (with predicate \texttt{below())}
They have the longest query execution times (up to four minutes for query \#18). 
The cause is twofold.
First, fetching the position of all elements requires some time.
Second, the comparison of source and target elements is done between all pairs of elements. This has a complexity in $O(n^2)$.
This significantly increases the computation time when many elements are compared to each other.

\subsubsection{Conclusion on Performance}

This experiment shows that there are differences in the execution speed of queries.
In a nutshell, basic queries run fast, visual queries (based on color and direction) are slow.
This is important for the developer of Abmash mashups: given these figures, she is able to tune her queries.
Abmash mashups are meant to be run in a batch manner (e.g. once a day). Even if some queries are slow, if the mashup execution takes a few hours, this may still be acceptable for a migration or a synchronization to be performed during the night.

However, some queries are really slow and may hinder the feasibility of running the mashup in a reasonable time.
This calls for optimizing the framework.
However, the primary design requirements is \emph{ease of use} and not performance. 
For instance, there is a caching mechanism which fetches additional information for each element in order to minimize the execution time for future lookups on these elements.
This caching is useless is some cases, and accounts for a significant part of the execution time of queries \#18-\#20.
Hard optimization is out of the scope of this paper.

We would like to conclude by saying that Abmash is liberal in its usage. 
On the one hand, the developer has powerful, intuitive Abmash selectors.
One the other hand, for performance, she can use and tune CSS, jQuery or xPath selectors \emph{in the same mashup}.

\subsection{Limitations of Abmash}
\label{discussion}

We conclude this  section by a qualitative discussion about the shortcomings and tradeoffs of Abmash.

An important contribution of Abmash is the element matching mechanism based on visual rendering.
This allows developers to write ``easy'' queries with no knowledge of the HTML document structure, as well as no knowledge of a DOM query language (such as XPath, CSS selectors and JQuery expressions).
Conversely, such well-known languages are more precise and come with much better tooling. For instance, using the web development tool Firebug, developers can obtain an XPath expression by a simple click on the element to be matched. There is a trade-off between the time spent to set up a query and its precision.
Abmash clearly balances towards rapid prototyping of queries.

Let us now discuss the lifecycle of Abmash integration programs.
Abmash programs are developed to integrate legacy web applications: being legacy they are well-known in one's company and their data and function semantics is integrated into the business habits.
Hence, the requirements of Abmash integration programs are relatively easy to be written (users know what they want to automate, they already do so manually).
As a result, Abmash integration programs are written in a short amount of time by small teams of developers.
Also, the stability of the programs being integrated has a direct impact on the stability of Abmash programs:
if the applications to be integrated do not change, Abmash programs are not susceptible to change either.
The biggest threats to the correctness of Abmash programs are look'n'feel changes: if the UI elements are resized or moved around, this can completely mix up the Abmash program.
The gravity depends whether Abmash actually finds an element or not. If Abmash finds no element where one is expected, it fails gracefully. If it finds the incorrect one, well, there is no way for the program to suspect it as wrong. It may then produce incorrect data. It is the responsibility of the engineer supervising the integration to check that the UI of the web applications under study do not change unexpectedly.

It is to be noted that Abmash programs completely rely on the built-in logic of the underlying applications being integrated.
If an integration program is run twice, it is of the responsibility  of the underlying applications to perform the necessary checks.
For instance, a target blogging app may -- or may not -- verify that a post with the same title already exists.

The current implementation of the Abmash framework is in Java. This choice was incidental, mainly driven by the fact that Selenium has a mature and stable Java API. This may also help Abmash developers with the advantages of static typing. However, the Abmash concepts and its fluent API are likely to also be implementable in other languages, especially scripting languages such as Javascript and Ruby.

We have loosely measured the performance of the Abmash prototype. First because our main goal was to improve the ease of writing integration programs, 
second, because our prototype was neither designed nor optimized with this point in mind. 
However, we have never experienced performance issues in a real setup (the first author uses the framework in his own business).

\section{Related Work}
\label{related_work}

Our work is at the confluence of different research areas:
mashups and end-user programming;
software maintenance and legacy code handling;
web data extraction.
It is out of the scope of this paper to extensively discuss the literature of all of those fields in full depth.
In the following, we thus draw a concise comparison with those domains.
Subsequently, we elaborate a bit more on mashup and web data extraction approaches. We often refer to the design requirements that we stated in section \ref{sec:req} when comparing Abmash with related work.

In short, a large part of contributions related to mashups focus on either end-user programming, or composing existing web services.
Abmash does not contribute to end-user programming;
the target user is rather a Java developer.
Furthermore, Abmash handles legacy web applications with no web service at all. 
Section \ref{rw-mashup} discusses this point in more details.

Much of the work on software maintenance focuses on migrating legacy applications to 
the web \cite{canfora2006migrating} and only a few papers discuss their integration. 
For instance Vinoski \cite{Vinoski2003} showed that integrating legacy code 
with classical middleware requires to invasively modify the code.
Sneed presented \cite{Sneed2003,Sneed2006} an approach to integrate legacy software into a service oriented architecture. 
It consists of automatically creating XML output from PL/I, COBOL and  C/C++ interfaces, 
which can be wrapped into a SOAP-based web service. 
Abmash neither modifies existing code nor ports applications to the web: it integrates legacy web applications.
There are also some pieces of research on automated data 
migration (e.g. \cite{drumm2007quickmig}). 
However, data migration is only one aspect of application integration, 
the scope of Abmash is larger.

Web data extraction research explores how to improve or automate the extraction of structured information from the web. 
Abmash does have a web data extraction aspect, but this is only a part of the contribution. 
We have shown in the paper that Abmash also provides means to navigate into a legacy web application, 
to submit data, and to expose legacy web applications as web services.
However, our approach has some roots in the vision-based extraction contributions, 
which are  discussed in \ref{rw-visual}.

\subsection{Mashups}
\label{rw-mashup}

According to the literature, the term 
"mashup" refers to programs that are related to the concepts of \emph{composition} and \emph{ease of development}.
For instance, Yahoo Pipes \cite{fagan2007mashing} composes RSS feeds to create new ones;
they are programed through an intuitive graphical user-interface.
The ease of development in mashup development is often associated 
to \emph{end-user programming} (e.g. \cite{wong2007marmite}), meaning that persons 
with little or even no programming education can still be able to create programs with an appropriate infrastructure.

Several authors \cite{Grammel2008,Gamble2008,Hartmann2008,koschmider2009elucidating,beemer2009mashups} 
have tried to set up elements of an orderly classification of the subject. 
In particular, there are \emph{data mashups} \cite{Gamble2008} (see \ref{sec:data-mashups}), \emph{process mashups} \cite{Grammel2008} (see \ref{sec:process-mashups}),  and \emph{presentation mashups} \cite{koschmider2009elucidating}  (see \ref{sec:presentation-mashups}).

Apart from those three main types of mashups, the literature also refers to ``enterprise mashups'' (e.g. \cite{Gamble2008}) for mashups created in an enterprise environment with some associated business value, and to ``mashup agents'' \cite{beemer2009mashups} for agents capable of semantically determining relevant information sources with respect to a specific concern. Finally, as others programming activities, mashup development 
is also related to development environments \cite{yu2007mixup,Grammel2010} and debugging \cite{cao2010debugging}.

\subsubsection{Data Mashups} 
\label{sec:data-mashups}
\emph{Data mashups} \cite{Gamble2008} (aka ``information mashups'' \cite{Grammel2008}) extract, filter, and combine  data from various sources (in possibly various formats).
The resulting information is meant to be human-readable (e.g. a web page) or machine processable (e.g. a RSS feed, a web service). Data mashups often require mashup frameworks encapsulating extraction routines and composition operators.

Damia is an IBM platform for creating and exposing data mashups \cite{altinel2007damia,simmen2008damia}.
\cite{Chang2005} presented  a system that integrates online databases in order to create meta-queries  over different databases. Meta-queries are a kind of data mashups. 
\cite{Tuchinda2008}'s system enables end-users to mash up data in a drag-and-drop based GUI called Karma. 
\cite{Lin2009}'s Vegemite is a data-centric mashup tool realized as Firefox plugin. It tracks the users' actions and stores selected data in tabular form.
\cite{Wang2009}'s Mashroom consists of joining data tables. An end-user interface uses spreadsheets to visualize the creation process.
\cite{wong2007marmite} presents an approach and a tool (Marmite) to extract data from different web sources in order to create information mashups. 
MashMaker is a similar concept as Marmite, while offering a more sophisticated script language to define Mashups \cite{ennals2007mashmaker} .
\cite{Hartmann2007} proposed d.mix to easily combine output of web pages in a programmable wiki-inspired application. It uses XPath and CSS selectors to fetch the elements and web APIs to manipulate them. 
Potluck is a data mashup tool for non-programmers \cite{huynh2008potluck}. 
which allows the easy creation of advanced visualizations of data, which can be further filtered across multiple dimensions. Potluck has a user-friendly interface which allows mashup creators to enter multiple web sources which are automatically analyzed. The contents from all sources can be arranged to create combined visualizations of interesting data.
Kongdenfha and colleagues \cite{Kongdenfha2009} explore the use of the spreadsheet paradigm for presenting data from the web.

Contrary to data mashups, our approach aims at \emph{interacting} with web applications,  which means retrieving data but also submitting data (R2). Also, previous work on data mashups does not support grabbing data that is only available with AJAX-based interactions (R3).

\subsubsection{Process Mashups}
\label{sec:process-mashups}

\emph{Process mashups} \cite{Grammel2008} (aka ``functionality mashups'' \cite{koschmider2009elucidating}), orchestrate different functionalities from different applications. A key point of process mashups is to handle the heterogeneity of interface and implementation technologies.

Process mashups are built on the idea of creating workflow orchestrating programmable interfaces.
Depending on the emphasis, the related work can be structured around those two axes.

\paragraph{Workflow-based Mashups}

\cite{koch2004integration} considered \textit{business processes} as a kind of end-user mashup since they are written by specialist end users. They proposed a bridge between business process models and web application models and presented a model-driven generative approach for realizing the mashups.

\cite{tony2005agentbased} presented an approach to wrap existing web services with standardized agents to integrate web services into workflow management systems. 
The processes are modeled by the use of colored Petri Nets, so that end users without programming skills are able to modify the definitions if the business requirements have changed. 

\cite{curbera2007bite}'s Bite  is a workflow-based script language to interact with REST interfaces, Java and JavaScript method invocations. These resources can be composed  mashups by using \textit{activities} and \textit{links}: ``Activities define units of work and links define dependencies between activities''. The Bite language focuses on simple generation of workflows with involved human interactions. These interactions range from real-world tasks like packaging a product for shipment to browser interactions like approving an order.

\cite{Pautasso2009} allows mashing up RESTful web services as a workflow mode in a tool called Joperal.
The control flow engine triggers multiple tasks automatically if they are a dependent of each other.
The transfer of data between the services needs is defined both for input and output
and the composition itself is configured by XML files.

With respect to our goal, workflow-based mashup techniques (and more generally model-based mashup techniques such as as MashArt \cite{Daniel2009}) are not appropriate to integrate legacy applications with no programmable interfaces (R1).

\paragraph{API-oriented Mashups}
According to \cite{maximilien2007domainspecific} mashups consist of three primary components: \emph{data mediation} components aggregate data from multiple services, \emph{process mediation} components orchestrate different services to create new processes and \emph{user interface} components create end-user interfaces to operate the combined services and visualize the results. 
\cite{hoyer2008design} described the generic design principles of API-oriented mashups. 
\cite{cappiello2009qualitymodel} state that mashup developers do not necessarily need advanced programming skills and are able to compose even complex mashups. Nevertheless they are bound to components that are developed to be accessible via standardized protocols and interfaces like REST or SOAP. The same arguments apply for ``Enterprise Mashup Application Platform'' \cite{Gurram2008} and \cite{Lopez2009}'s approach .
In all these approaches, they define mashup as an application that only uses existing APIs.

Our approach does not make this strong assumption, we can mashup any kind of web applications, even in the absence of programmable interfaces (R1).

We note that Bgu et al.'s approach \cite{ngu2010semantic} \emph{enables progressive composition of non Web service based components}, however it requires semantic annotations for composing the components together. Such annotations are not API \emph{per se} but are conceptually close.
Similarly, Oren et al's discuss \cite{oren2007flexible} the notion of mashing-up in the semantic web which is a kind of universal data-oriented API.

\subsubsection{Presentation Mashups}
\label{sec:presentation-mashups}

\emph{Presentation mashups} offer a new user interface to browse and manipulate some existing data (also referred to as ``web page customizations'' \cite{Grammel2008}). 
Different kinds of mashups can be stacked, for instance, a presentation mashup may present the results of a data mashup.
 
Some mashups approaches consist of enabling users to customize the user interface of web applications.
For instance, {\em Mixup} \cite{yu2007mixup}' is a mashup framework to combine user interfaces 
as an intelligent union of elementary UIs. 
{\em Exhibit} is a framework that enables individuals to provide rich web-interfaces for 
browsing data  \cite{Huynh2007}.
\cite{hornung2008mashupsdeepweb} developed a browser extension capable of interacting with forms, collecting and composing data from different web pages, and presenting the queried information. The interaction scripts depend on the document's source code.
In comparison, Abmash focuses on integration mashup, which is much more a kind of data and process mashup.

Chickenfoot is an automation framework that aims at creating mashups in the sense of automating repetitive operations, and integrating multiple websites \cite{bolin2005chickenfoot}.
It offers intuitive methods to find and interact with web page elements. This is done by pattern matching algorithms to find elements through their visible labels. 
However, Chickenfoot only offers  very basic functionality to extract content from web pages. It is capable of returning the visible text and source code of specific elements, but more complex information like geographic locations or tabular data can not be extracted (contrary to Abmash). In other words, Abmash is conceptually close to Chickenfoot, but goes further with respect to visual semantics and automated data extraction.

\cite{little2007koala} and \cite{leshed2008coscripter} introduced CoScripter/Koala for enabling end users to record process in a readable and editable format based on natural language. Their execution engine also uses the visual semantics of web pages to execute the scripts. The main difference with our approach is that writing the script in Java enables a full range of complex data transformation and manipulation that is not possible with this CoScripter/Koala.

\subsubsection{Mashup Tools and Libraries}

A mashup is basically a assembly of data and functions of web applications.
Hence, any tool that is able to create HTTP connections and extract some information from web pages could be used for creating mashups.

There are several libraries like \textit{lxml}\footnote{lxml: \url{http://lxml.de/} \urlaccessed}, \textit{Mechanize}\footnote{Mechanize: \url{http://wwwsearch.sourceforge.net/mechanize/} \urlaccessed} or \textit{Splinter}\footnote{Splinter: \url{http://splinter.cobrateam.info/} \urlaccessed} that support parsing web pages and interacting with web applications.
They all require in-depth knowledge about the technical source code of web pages, contrary to Abmash.

\textit{Sahi}\footnote{Sahi: \url{http://sahi.co.in/w/sahi} \urlaccessed}  offers a scripting language 
to find page elements and to interact with web applications for testing purposes.
Finding those elements can be achieved by including visual attributes like visible texts in
the querying process, e.g., \texttt{\_link("Link to Sahi website")}. 
It also introduces the functions \texttt{\_near} and \texttt{\_in}, which consider geometrical 
relations between elements of the document structure. 
Sahi goes in the same direction as Abmash, but in a very limited way (no handling of visual semantics based on elements and font size, no handling of tabular data, etc.).

The \textit{Yahoo! Query Language} (YQL) is an SQL-like query language, which allows queries for  data across popular web services throughout the internet \footnote{YQL: \url{http://developer.yahoo.com/yql/} \urlaccessed}. Different virtual database tables represent the various types of data available. 
The YQL is along the same line as \cite{Huy2005} who proposed to wrap web sites into web services to can be called by client integration programs.

\subsection{Web Data Extraction}
\label{rw-visual} 

Web data extraction research explores how to improve or automate the extraction of structured information  from the web.
In this area, there is a line of research on data extraction with visual information.

Lixto \cite{baumgartner2001visual,baumgartner2003visual} is a web information extraction system where users visually select elements to be extracted on the screen.
Lixto then infers an matching expression using an internal datalog language called Elog.
While Lixto and Abmash share the ``visual'' focus, they are actually different: Lixto is about an interactive graphical user-interface,
Abmash focuses on predicates on the visual rendering of web pages.
Similarly, \cite{jan2005wise}'s graphical interface enables users to manually select items. 
Those items are then used to train a system in order to extract them automatically.
Compared to Abmash, the proposed system targets end-user programming, hence trades the possibility of writing complex integration applications for simplicity.The work we present here focuses on a different trade-off: programmers write complex  integration applications without comprehending the internals of the applications to be integrated.

VIPS \cite{cai2003extracting,deng2003vips} is a page segmentation algorithm which aims at distinguishing between multiple parts of web pages. Each of these parts have a different semantic value and is divided into a hierarchy of visible blocks, such as headers, footers or text paragraphs. VIPS simulates how a user understands a web page by the examination of the hierarchy of page blocks, the detection of visual separators and the analysis of the content structure. 
Song et al. \cite{song2004blockimportance} use the VIPS approach to dive further into the distinction of different blocks and add learning methods to create a ``block importance model''. They use neural network learning to rate the importance of each block. Furthermore, Support vector machines are used to classify the different blocks of a web page. 
 \cite{simon2005viper,liu2006vision,liu2010vide} also use the visual rendering to extract repetitive patterns in web pages such as search engine results.

Those pieces of research on the visual semantics only focus on extracting data, 
they do not provide the possibility to interact with web pages (R2) possibly with AJAX/Javascript (R3). 
Hence, our contribution goes further: we use some visual information to drive the interaction 
(data extraction and submission) with legacy web applications.

\section{Conclusion}
\label{conclusion}

In this paper, we have presented a framework called Abmash to write ``integration mashups'', that are short programs to interact with legacy web applications.
The key insight of Abmash is that integration mashups can be written by programmatically imitating human understanding and interactions with web applications. As a result, integration mashups can be written without invasive software development, i.e., without hooking into legacy code. At the same time, such human interactions based code benefits from the logic that is given to human users to prevent incorrect data and behavior. 

We have presented integration mashups for data migration and API creation, but the framework is meant to liberate the creativity of programmers and many other integration mashups can be envisioned using the framework (e.g. UI customization).
Beyond creativity, Abmash aims at bringing economic value by lowering the cost of writing integration code. The evaluation actually presented first pieces of evidence that developers can easily write correct integration programs (short programs written in a short amount of time).

The whole framework is released as open-source project\footnote{Abmash at Github: \url{http://github.com/alp82/abmash} \urlaccessed} to create a community-driven and mature web automation tool. It is used by the first author in his own business. Future work will focus on integrating Abmash with improved visual queries as well as workflow paradigms and orchestration technologies.

\bibliographystyle{wileyj} 
\bibliography{thesis}

\end{document}